\documentclass[11pt]{article}
\usepackage{graphicx}

\newcommand{\BABARPubYear}    {04}

\newcommand{\BABARConfNumber} {03}
\newcommand{\SLACPubNumber} {10622}
\newcommand{\LANLNumber} {0000}

\input babarsym
\newcommand{\Denu}{\ensuremath{D\xspace\electron\nub}}
\newcommand{\shmax}{\ensuremath{s_{\mathrm{h}}^{\mathrm{max}}}}

\newcommand{\GEVCC}{\ensuremath{{\mathrm{\,Ge\kern -0.1em V^2\!/}c^4}}\xspace}

\newcommand{\VVub}{4.57}
\newcommand{\VEstatVub}{0.21}
\newcommand{\VEdetVub}{0.25}
\newcommand{\VEbkgVub}{0.34}
\newcommand{\VEfuVubp}{+0.59}
\newcommand{\VEfuVubm}{-0.29}
\newcommand{\VEhqeVub}{0.22}
\newcommand{\VVubBelle}{4.99}
\newcommand{\VEstatVubBelle}{0.23}
\newcommand{\VEdetVubBelle}{0.25}
\newcommand{\VEbkgVubBelle}{0.34}
\newcommand{\VEbuVubpBelle}{+0.04}
\newcommand{\VEbuVubmBelle}{-0.18}
\newcommand{\VEfuVubpBelle}{+0.18}
\newcommand{\VEfuVubmBelle}{-0.14}
\newcommand{\VEhqeVubBelle}{0.22}
\newcommand{\VBF}{2.37}
\newcommand{\VEstatBF}{0.22}
\newcommand{\VEdetBF}{0.26}
\newcommand{\VEbkgBF}{0.34}
\newcommand{\VEfuBFp}{+0.60}
\newcommand{\VEfuBFm}{-0.30}
\newcommand{\VUBF}{4.51}
\newcommand{\VEstatUBF}{0.42}
\newcommand{\VEdetUBF}{0.50}
\newcommand{\VEbkgUBF}{0.66}
\newcommand{\VEfuUBFav}{0.19}
\newcommand{\VBFBelle}{2.76}
\newcommand{\VEstatBFBelle}{0.26}
\newcommand{\VEdetBFBelle}{0.30}
\newcommand{\VEbkgBFBelle}{0.40}
\newcommand{\VEbuBFpBelle}{+0.05}
\newcommand{\VEbuBFmBelle}{-0.20}
\newcommand{\VEfuBFpBelle}{+0.20}
\newcommand{\VEfuBFmBelle}{-0.16}
\newcommand{\VUBFBelle}{4.46}
\newcommand{\VEstatUBFBelle}{0.42}
\newcommand{\VEdetUBFBelle}{0.49}
\newcommand{\VEbkgUBFBelle}{0.64}
\newcommand{\VEfuUBFavBelle}{0.19}
\newcommand{\btouenu}{\ensuremath{\b\to\u\electron\nub}}
\newcommand{\btocenu}{\ensuremath{\b\to\c\electron\nub}}
\newcommand{\mbSF}{\ensuremath{m_b^{\mathrm{SF}}}}

\long\def\inst#1{\par\nobreak\kern 4pt\nobreak
    {\it #1}\par\vskip 10pt plus 3pt minus 3pt}

\begin{document}
{\pagestyle{empty}
\setcounter{footnote}{0}

BaBar analysis document 846, version 13.
\begin{flushright}
\babar-CONF-\BABARPubYear/\BABARConfNumber \\
SLAC-PUB-\SLACPubNumber \\
hep-ex/\LANLNumber \\
August 2004 \\
\end{flushright}

\par\vskip 5cm

\begin{center}
\Large \bf Determination of \Vub\ in inclusive semileptonic \B\ meson decays
\end{center}
\bigskip

\begin{center}
\large The \babar\ Collaboration\\
\mbox{ }\\
\today
\end{center}
\bigskip \bigskip

\begin{center}
\large \bf Abstract
\end{center}
We present a preliminary determination of the CKM matrix element \Vub\ based on
the analysis of semileptonic \B\ decays from a sample of $88$ million
\FourS\ decays collected with the \babar\ detector at the 
\pep2\ \epem\ storage ring.  Charmless semileptonic \B\ decays 
are selected using the electron energy $E_e$ and the invariant
mass $q^2$ of the electron-neutrino pair.  The neutrino momentum is
inferred from a measurement of the visible energy and momentum in the
detector and knowledge of the \epem\ beam momenta.  The partial
branching fraction is determined in a region of the $q^2$-$E_e$
plane where semileptonic \B\ decays to charm 
are highly suppressed.
The total charmless semileptonic branching fraction is
extracted using a theoretical calculation based on the heavy quark
expansion.
Preliminary results yield 
$\Vub = (\VVub\pm \VEstatVub\pm \VEdetVub\pm \VEbkgVub
~{}^{\VEfuVubp}_{\VEfuVubm} \pm \VEhqeVub)\times 10^{-3}$ 
where the uncertainties are from statistics (data and MC), detector
modeling, background modeling, the shape function, and the heavy quark
operator product expansion, respectively.

\vfill
\begin{center}

Submitted to the 32$^{\rm nd}$ International Conference on High-Energy Physics, ICHEP 04,\\
16 August---22 August 2004, Beijing, China

\end{center}

\vspace{1.0cm}
\begin{center}
{\em Stanford Linear Accelerator Center, Stanford University, 
Stanford, CA 94309} \\ \vspace{0.1cm}\hrule\vspace{0.1cm}
Work supported in part by Department of Energy contract DE-AC03-76SF00515.
\end{center}

\newpage
} 

%
%
\begin{center}
\small

The \babar\ Collaboration,
\bigskip

%
B.~Aubert,
R.~Barate,
D.~Boutigny,
F.~Couderc,
J.-M.~Gaillard,
A.~Hicheur,
Y.~Karyotakis,
J.~P.~Lees,
V.~Tisserand,
A.~Zghiche
\inst{Laboratoire de Physique des Particules, F-74941 Annecy-le-Vieux, France }
% ...

\end{center}\newpage

\section{INTRODUCTION}
\label{sec:Introduction}
The study of the weak interactions of quarks has played a crucial role in the
development of the Standard Model (SM), which embodies our understanding of
the fundamental interactions.  The increasingly precise measurements
of CP-violating asymmetries in \B\ decays allow stringent experimental
tests of the SM mechanism for CP violation~\cite{ref:KM}
via the non-trivial phase in the Cabibbo-Kobayashi-Maskawa (CKM) matrix.
Improved determinations of \Vub, the coupling strength of the \b\ quark to the
\u\ quark, will improve the sensitivity of experimental tests of
the SM description of CP violation.

Early determinations~\cite{ref:firstVub} of \Vub\ were based on the
study of the lepton momentum spectrum near the kinematic endpoint in
semileptonic \B\ decays, where the background from the dominant decay
chain\footnote{ Throughout this paper, whenever a mode is given, the
charge conjugate is also implied.}  $\btocenu$ is
kinematically forbidden.  
See Ref.~\cite{ref:endpoint} for more recent measurements using this method.
This technique provides precise measurements
of the partial branching fraction for lepton
momenta\footnote{Throughout this paper, quantities are given in the
\FourS\ rest frame unless stated otherwise.}  above $\sim 2.3\,\gevc$,
but accepts only $\sim 10\%$ of the $\btouenu$ rate, resulting
in a substantial theoretical uncertainty in the determination of \Vub.
The major part of this uncertainty comes from limited knowledge of the
``shape function'' (SF), i.e.~the distribution of the \b\ quark
momentum inside the \B\ meson~\cite{ref:SF}. These uncertainties can be
reduced by measuring the energy spectrum of photons from the
decay $\b\to\s\gamma$~\cite{ref:SF,ref:bsgSF}, but this is very 
challenging~\cite{ref:bsgSpectrum}.

Recently, measurements of \Vub\ that use the invariant mass $m_X$
of the hadronic system in semileptonic \B\ decays have
appeared~\cite{ref:mhVubBaBar,ref:mhVubBelle}.  The use of $m_X$
to select $\btouenu$ decays results in a much higher 
acceptance ($\sim 70\%$) of the decay rate.  It is, however,
experimentally challenging, since the
association of particles with the semileptonic \B\ decay is rendered
difficult by the presence of the decay products of the $\Bbar$ in the
event.  These measurements also have significant SF uncertainties.

In this analysis a new approach~\cite{ref:shmax} is taken to the
determination of \Vub.  
Semileptonic \B\ decays are selected using energetic electrons and 
simultaneously making requirements on $q^2$,
the invariant mass squared of the $\electron\nub$ pair.  
The neutrino 4-momentum is reconstructed from the visible 4-momentum and knowledge
of the \epem\ initial state.
The dominant charm background is then suppressed by selecting a region of
the $q^2-E_{e}$ phase space where properly reconstructed $\btocenu$ 
events are kinematically excluded.
The fraction of the $\btouenu$ phase space accepted is $\sim 20\%$.
The amount of background events remaining in the signal region due to resolution 
effects is evaluated in Monte Carlo simulations.  
The determination of \Vub\ in this method is sensitive to the \b\ quark mass
through both the heavy quark operator product expansion~\cite{ref:HQE} 
(HQE) and the SF, but has little sensitivity to Fermi motion
as described below.

\section{DATASET AND SIMULATION}
\label{sec:babar}
The data used in this analysis were collected with the \babar\
detector at the \pep2\ asymmetric-energy \epem\ storage ring between
2000 and 2002.  
The \babar\ detector is described in detail
elsewhere~\cite{ref:babar}.  
A sample of $81.4\invfb$ collected at the \FourS\ resonance, 
corresponding to $88.4$ million $\BB$ pairs, is used along
with $9.6\invfb$ collected at center-of-mass energies approximately
$40\,\mev$ below 
\BB\ threshold.  The data below \BB\ threshold, 
scaled in cross-section and luminosity, and whose particles are
scaled in energy, are used to subtract the
non-\BB\ contributions from the data collected on the \FourS\ resonance.
Simulated \BB\ events are used in estimating efficiencies and
backgrounds.  Branching fractions and form factors are taken 
from Ref.~\cite{ref:pdg2002} in most cases.  Branching fractions for the
semileptonic \B\ decays to charm are adjusted as described below.  The
simulation of charmless semileptonic \B\ decays is based on the model
described in Ref.~\cite{ref:NdF}, which calculates the triple differential decay
rate to order $\alpha_S$ and convolutes it with a shape function parameterized as
$$
F(k_{+}) = N (1 - x)^{a}\ e^{(1+a)x};
\hspace{1cm}x\equiv \frac{k_{+}}{m_B-\mbSF}\le 1,
$$
where $k_{+}$ is the residual $b$-quark momentum, to account for the non-perturbative
interactions of the $b$ quark within the \B\ meson.
We use~\cite{ref:Gibbons} $\mbSF=4.735\,\gevcc$ and $a=1.6$.  
This model produces a spectrum of hadronic masses $m_X$ down 
to $2m_\pi$ that does not contain any resonant states.  
Subsequent fragmentation of the mesons was simulated 
via JETSET~\cite{ref:SJostrand}.  Decays to low-mass 
hadrons ($\pi$, $\eta$, $\rho$, $\omega$, $\eta^{\prime}$) are simulated
separately, using the form factor model of Ref.~\cite{ref:ISGW2}, 
and mixed with the non-resonant states in such a way as to
keep the $m_X$, $q^2$ and $E_e$ spectral distributions the same as in the 
inclusive model. 

\section{ANALYSIS METHOD}
\label{sec:Analysis}
\begin{figure}[!htb]
\begin{center}
\includegraphics[height=6cm,clip=]{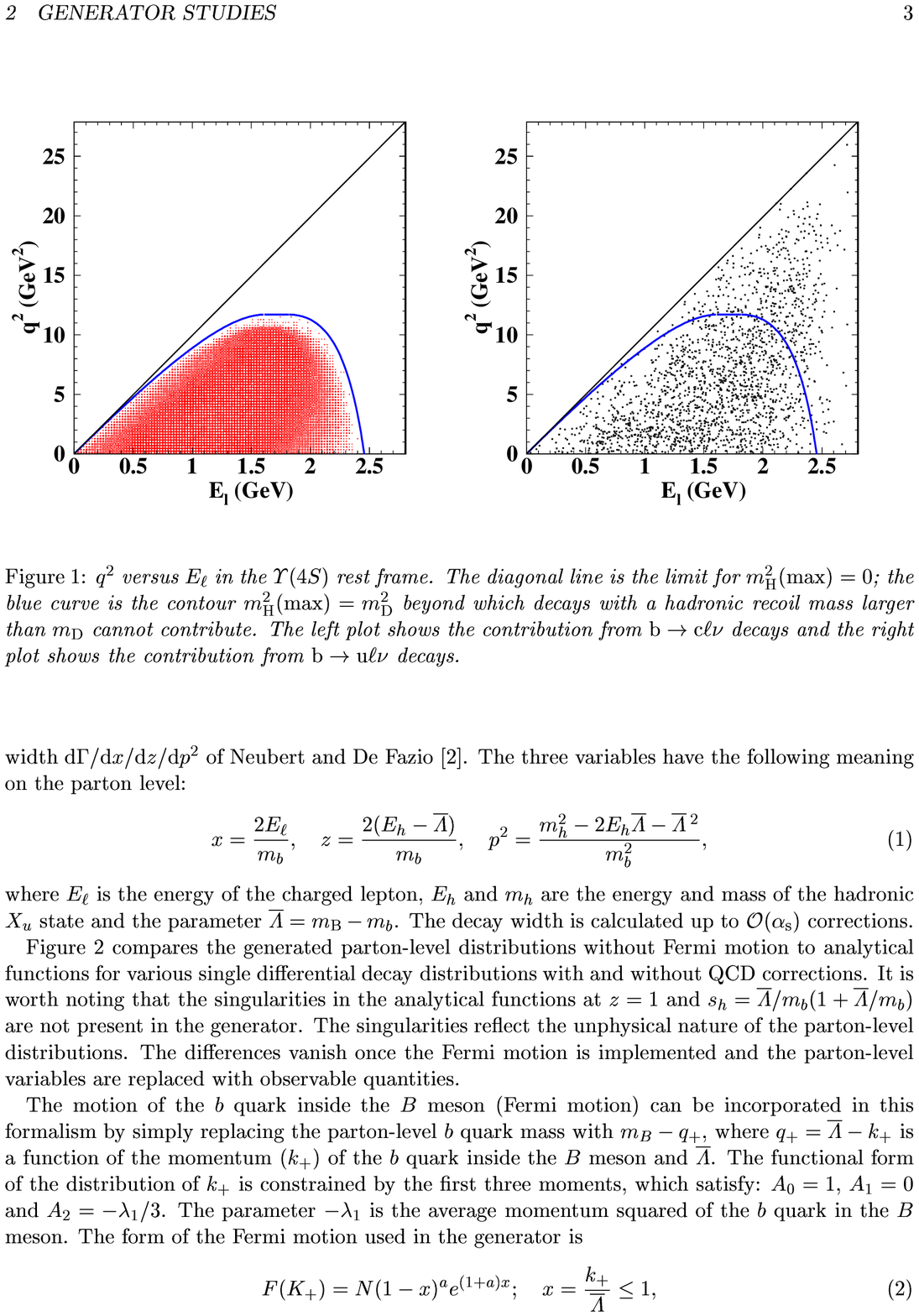}\\
\caption{Distribution of $q^2$ versus $E_e$ for generated \btocenu\ (left)
and \btouenu\ (right) events in the \FourS\ rest frame.  The 
curved contour corresponds to $\shmax=m_D^2$ and the diagonal line 
to $\shmax=0\,\GEVCC$.
}
\label{fig:q2Ee}
\end{center}
\end{figure}

Events are selected based on the presence of an identified electron
with energy $E_e>2\,\gev$ using the criteria given
in Ref.~\cite{ref:elecID}.  Electrons from the decay $\jpsi\to\epem$ are
vetoed.  The criteria given in Ref.~\cite{ref:elecID} for selecting
hadronic events and rejecting radiative Bhabha events are applied.
Within these events the total visible 4-momentum $p_\mathrm{vis}$ is
determined using charged tracks emanating from the collision point,
identified pairs of charged tracks from $\KS\to\pip\pim$,
$\Lambda\to\proton\pim$ and $\gamma\to\epem$, and energy deposits in
the electromagnetic calorimeter.  Each charged particle is assigned 
a mass hypothesis based on particle identification information.
Only those calorimeter clusters
unassociated with a charged track and which have a lateral energy
spread consistent with photons are considered.  Energy deposits due to
long-lived neutral hadrons are not efficiently rejected by these
criteria and are treated as coming from massless particles in the
4-vector sum.
We form the missing 4-momentum
$p_{\mathrm{miss}}=p_{\epem}-p_{\mathrm{vis}}$, where $p_{\epem}$ is
the 4-vector of the initial state.  
The 4-vector $ p_{\nu} = ({\bf P}_{\nu},|{\bf P}_{\nu}|) $ is used
as an estimate of the momentum of the neutrino from the decay $\btouenu$,
where ${\bf P}_\nu$ is derived from ${\bf P}_{\mathrm{miss}}$ by
applying a 
bias correction that was determined from Monte Carlo signal events.
Additional requirements are made to improve the quality of the
neutrino reconstruction and suppress contributions from
$\epem\to\qqbar$ continuum events.  Each event must satisfy (1) no
additional identified leptons, (2)
$-0.95<\cos\theta_{\mathrm{miss}}<0.8$, (3)
$0.0<E_{\mathrm{miss}}-|{\bf P}_{\mathrm{miss}}|c<0.8\,\gev$, (4)
$0.0<|{\bf P}_\mathrm{miss}|<2.5\,\gevc$ and (5)
$|{\bf P}_e\cdot {\bf T}|<0.75\,|{\bf P}_e|\,|{\bf T}|$, where
$\theta_{\mathrm{miss}}$ is the angle of the missing momentum vector
with respect to the beam axis, and ${\bf T}$ is the thrust vector for 
the event, excluding the signal electron candidate.  
The large background from
$\btocenu$ decays is then suppressed by calculating $q^2=(p_e+p_{\nu})^2$ and
computing the maximum kinematically allowed hadronic mass squared,
$\shmax$, for a given $E_e$ and $q^2$.  In the case where $\pm E_{e} >
\pm \frac{\sqrt{q^2}}{2}\left(\frac{1\pm\beta}{1\mp\beta}\right)$, the maximum
invariant hadronic mass squared is
\begin{eqnarray}
\shmax &=& m_{B}^2 +q^2 - 2m_{B}E_{e}\sqrt{\frac{1 \mp \beta}{1 \pm\beta}} 
- 2 m_{B}\left(\frac{q^2}{4 E_{e}}\right)\sqrt{\frac{1 \pm \beta}{1 \mp \beta}},
\hspace{3cm}
\\ 
{\mathrm{otherwise\hspace{2.25cm}}}& & \nonumber \\
\shmax&=& m_{B}^2 + q^{2} -2 m_{B}\sqrt{q^2}, \nonumber
\label{equation-shmax2}
\end{eqnarray}
where $\beta=0.06$ is the boost of the \B\ meson in the \FourS\ frame.
We require $\shmax< 3.5\,\GEVCC\simeq m_D^2$.
Figure~\ref{fig:q2Ee} shows the distribution of generated \btocenu\ and
\btouenu\ decays in the $q^2$-$E_e$ plane and indicates the contour
corresponding to $\shmax=m_D^2$.  The requirements for $E_e$ and $\shmax$ as well as 
criteria (1)--(5) were optimized to minimize the total (experimental and theoretical) 
uncertainty on $\mathcal{B}(\btouenu)$.  

\begin{figure}[!htb]
\begin{center}
\includegraphics[height=12cm]{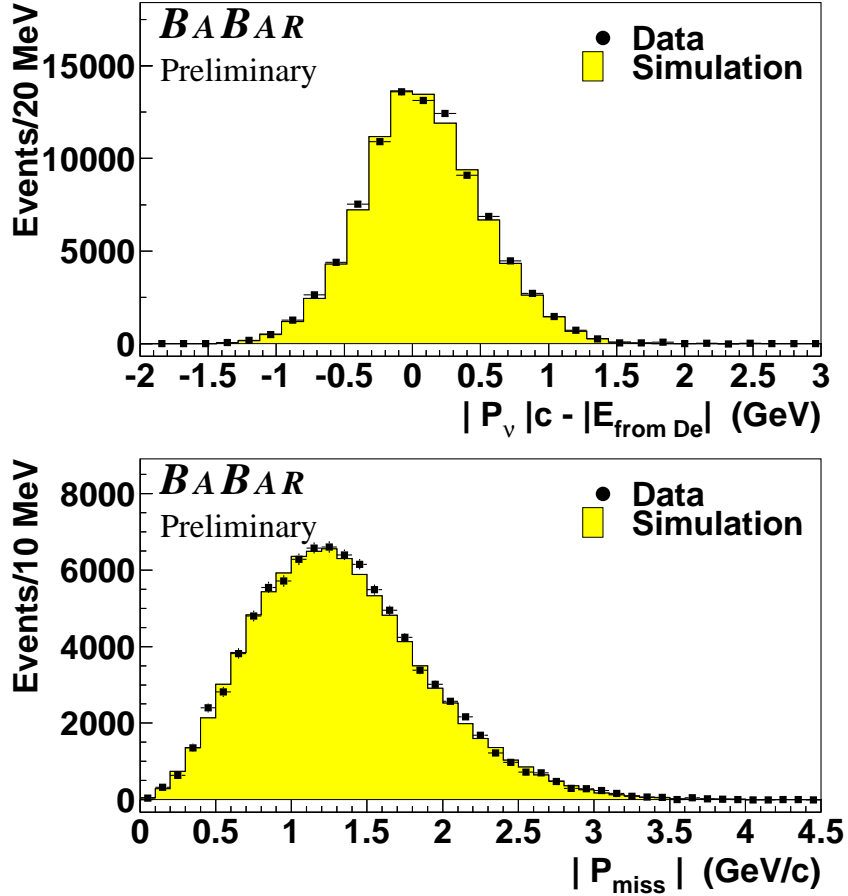}\\
\caption{The top plot shows the difference between the two neutrino 
energy estimates described in the text for continuum-subtracted data 
and $\BB$ MC. The bottom plot shows the $|{\bf P}_{\mathrm{miss}}|$ 
distribution for the \Denu\ control sample.}
\label{fig:nures}
\end{center}
\end{figure}

The quality of the neutrino reconstruction was evaluated using a
control sample (\Denu) consisting of $\sim 90,000$ decays of the type
$\Bbar\to\Dz\electron\nub(X)$ where the $\Dz$ is reconstructed in the
$\Km\pip$ decay mode and satisfies $|{\bf P}_{\Dz}|>0.5\,\gevc$ and
the electron satisfies $E_e>1.4\,\gev$.  The $\Dz$-$e$ combination
must satisfy $-2.5<\cos\theta_{B-De}<1.1$ where
$\cos\theta_{B-De}=(2E_B E_{De}-m_B^2-m_{De}^2)/(2|{\bf P}_B||{\bf P}_{De}|)$ is the 
cosine of the angle between the vector momenta
of the $B$ and the $\Dz$-$e$ system under the assumption that the only
missing particle in the $B$ decay is a single neutrino.  This
criterion selects both $\Bbar\to \Dz e\overline{\nu}$ and $\Bbar\to~\Dstar
\electron\nub,~\Dstar\to\Dz (\pi,\ \gamma)$
decays with high efficiency while rejecting
other sources of $\Dz$-$e$ combinations.  The additional 
requirements made on the semileptonic \B\ decay in the 
control sample selection differ from the signal selection, 
where only the electron is required.  However, in
neither case are any requirements made on the decay of the other \B,
leaving its properties the same in both samples.  
An estimate of the neutrino
energy can be formed from the known $B$ energy and the measured $\Dz$
and electron energies.
A second estimate of the neutrino energy is
given by the $|{\bf P}_{\nu}|$ defined above.  In simulation the first
(second) estimate for the true neutrino energy has a bias of $0.014\
(0.271)\,\gev$ and an r.m.s. of $0.202\ (0.359)\,\gev$ for \Denu\ events.  
Subtracting the first estimate from
the second gives the distribution shown in Figure~\ref{fig:nures},
from which we find a mean (sigma) of $0.109\ (0.433)\,\gev$ on data
and $0.107\ (0.427)\,\gev$ on MC, indicating that the missing energy
from the other \B\ decay is well modeled.  The distribution of
$|{\bf P}_{\mathrm{miss}}|$, also shown in Figure~\ref{fig:nures}, 
is sensitive to both the modeling of the semileptonic \btocenu\ 
decays and of missing energy, and shows good agreement with the simulation.

The \Denu\ control sample was also used to 
improve the modeling of the $\btocenu$ decays as follows.  
A binned $\chi^2$ fit in the variables $|{\bf P}_{D}|$, $E_e$ and
$\cos\theta_{B-De}$~\footnote{The $\cos\theta_{B-De}$ requirements 
were relaxed to $-10<\cos\theta_{B-De}<5$.}
is performed on the $\Denu$ sample after subtracting continuum and 
combinatorial background.
The fit determines scale factors for the MC components
$\Bbar\to De\nub$, $\Bbar\to \Dstar e\nub$ and other contributions,
of which 85\%\ are $\Bbar\to D^{**}e\nub$ (P-wave charm mesons) 
or $\Bbar\to D^{(*)}{\pi}e\nub$; the remainder comes 
from $\Bbar\to D\Dbar$ decays and from misidentified electrons.  
Updated branching
fractions are obtained by requiring the total $\btocenu$ branching
fraction to equal the measured value~\cite{ref:Elmoments}.  The fit
raises the branching fractions relative to those of 
Ref.~\cite{ref:pdg2002} for
$\Bbar\to De\nub$ and $\Bbar\to \Dstar e\nub$ and lowers the remaining
contributions.  Given the large uncertainty in the composition of the
high mass charm states in semileptonic \B\ decay, this procedure
cannot be considered a measurement of the branching fractions, but
rather a means of improving the simulation of inclusive quantities in
\btocenu\ decays.  We find an improved agreement between data and MC
in kinematic distributions ($E_e$, ${\bf P}_{\mathrm{miss}}$, $q^2$, $\shmax$) in the
inclusive electron sample using the fitted branching fractions.  These
revised branching fractions are used to determine the \btocenu\
background in the inclusive electron sample.  

The signal region
$E_e>2\,\gev$ and $\shmax<3.5\,\GEVCC$ was not examined until all
selection criteria were fixed based on studies of simulated data and
of control samples.  Two control samples are used to reduce the
sensitivity of the efficiency and background estimates to details of
the simulation: (a) the \Denu\ control sample described above; and (b)
events satisfying the normal selection criteria but having
$\shmax>4.25\,\GEVCC$ (high-\shmax\ sideband), a sample dominated by
background.  Efficiencies can be calculated,
separately in data and MC, from the ratio of \Denu\ candidates 
satisfying cuts (1)-(5) to the total number of 
selected \Denu\ candidates having $E_e>2.0\,\gev$.  The sensitivity
of the $\btouenu$ signal efficiency to details of the simulation is
reduced by multiplying it by the data/simulation ratio of 
\Denu\ efficiencies.
The high-$\shmax$ sideband is used to normalize the Monte Carlo $\shmax$
distribution to the data, reducing sensitivity to background
normalization uncertainties.  
The total charmless semileptonic branching fraction is calculated as
\begin{equation}
\mathcal{B}(\btouenu) = \frac{N^{\mathrm{data}}_{\mathrm{cand}}\ -\
M^{\mathrm{MC}}_{\mathrm{bkg}} 
\frac{N^{\mathrm{data}}_{\mathrm{side}}}{M^{\mathrm{MC}}_{\mathrm{side}}}}{
 2\ \epsilon^{\mathrm{MC}}_{u}\frac{\epsilon_{\Denu}^{\mathrm{data}}}
{\epsilon_{\Denu}^{\mathrm{MC}}} \ N_{\BB}},
\label{eq:bf}
\end{equation}
where $N^{\mathrm{data}}_{\mathrm{cand}}$ and
$N^{\mathrm{data}}_{\mathrm{side}}$ refer to the number of candidates
in the signal and high-$\shmax$ sideband regions of the data after 
subtraction of non-\BB\ contributions determined on data taken below
the \BB\ threshold,
$M^{\mathrm{MC}}_{\mathrm{bkg}}$ and $M^{\mathrm{MC}}_{\mathrm{side}}$
refer to background in the signal region and the yield in the sideband
region in simulated events, 
$\epsilon_{\Denu}^{\mathrm{data}}$ and $\epsilon_{\Denu}^{\mathrm{MC}}$ 
are the efficiencies calculated on the corresponding \Denu\ samples,
pseudo-efficiency of the event selection determined from the data and
Monte Carlo \Denu\ control sample events, 
$N_{\BB}$ is the number of $\FourS\to\BB$ decays analyzed.
The total efficiency times acceptance for $\btouenu$ decays in the 
simulation is given by 
$\epsilon^{\mathrm{MC}}_u = \epsilon_{\,\mathrm{sig}}f_u + \epsilon_{\,\overline{\mathrm{sig}}}(1-f_u)$,
where $f_u$ is the fraction of \btouenu\ decays generated in the signal
region, and $\epsilon_{\,\mathrm{sig}}$ ($\epsilon_{\,\overline{\mathrm{sig}}}$)
is the efficiency for an event inside (outside) the signal region to be
reconstructed and pass our selection requirements.  

The effects of detector response and the boost of the \B\ meson in the 
\FourS\ frame are unfolded to produce a partial branching fraction that 
can be compared directly to theoretical calculations.  We define
%
%
\begin{eqnarray}
\Delta\mathcal{B}(\btouenu) & \equiv & \mathcal{B}({\btouenu}) f_u \\ \nonumber
\label{eq:upbf}
&=& \frac{N^{\mathrm{data}}_{\mathrm{cand}}\ -\
   M^{\mathrm{MC}}_{\mathrm{bkg}} 
   \frac{N^{\mathrm{data}}_{\mathrm{side}}}{M^{\mathrm{MC}}_{\mathrm{side}}}}{
   2\ \frac{\epsilon_{\Denu}^{\mathrm{data}}}{\epsilon_{\Denu}^{\mathrm{MC}}}
    \ N_{\BB}}\ \frac{f_u}{(\epsilon_{\,\mathrm{sig}}f_u + \epsilon_{\,\overline{\mathrm{sig}}}
(1-f_u))}\\
&=& \frac{N^{\mathrm{data}}_{\mathrm{cand}}\ -\
   M^{\mathrm{MC}}_{\mathrm{bkg}} 
   \frac{N^{\mathrm{data}}_{\mathrm{side}}}{M^{\mathrm{MC}}_{\mathrm{side}}}}{
   2\ \epsilon_{\,\mathrm{sig}} \frac{\epsilon_{\Denu}^{\mathrm{data}}}{\epsilon_{\Denu}^{\mathrm{MC}}}
    \ N_{\BB}}\ \left[ 1 + \left( \frac{1}{f_u}-1\right) \frac{\epsilon_{\,\overline{\mathrm{sig}}}}{\epsilon_{\,\mathrm{sig}}}\right]^{-1}
.\\ \nonumber
\end{eqnarray}
%
Whereas the extraction of the total branching fraction $\mathcal{B}(\btouenu)$ has 
a strong dependence on the signal modeling due to the theoretical uncertainty 
on $f_u$, this dependence is suppressed in computing the unfolded partial branching 
fraction $\Delta\mathcal{B}(\btouenu)$ since the ratio 
$\epsilon_{\,\overline{\mathrm{sig}}}/\epsilon_{\,\mathrm{sig}} \approx 0.029$ 
is much smaller than 1 (see Table~\ref{tab:yields}).
%
 
\begin{figure}[!htb]
\begin{center}
\includegraphics[height=12cm]{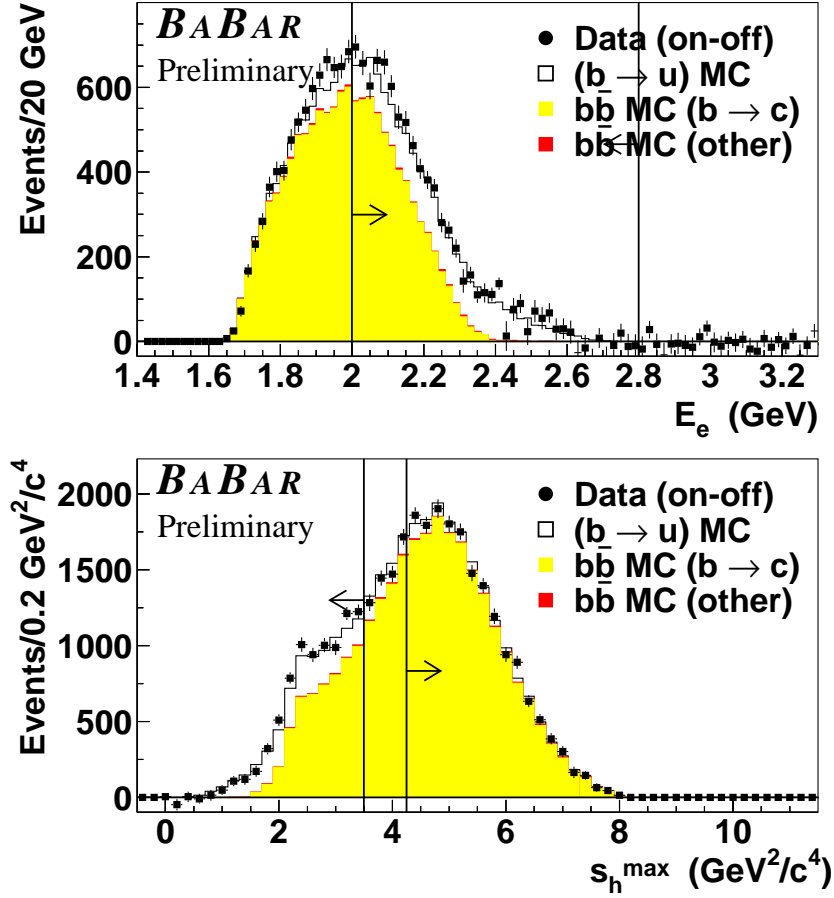}\\
\caption{The electron energy distribution (top) and $\shmax$
distribution (bottom) after cuts have been applied to all variables
except the one being plotted.  The arrows denote the signal region and also the
high-$\shmax$ sideband region (above $4.25\gev^2/c^4$).  The number of
background events from cascade decays and mis-identified electrons
is small (events denoted as other).}
\label{fig:plepstar}
\end{center}
\end{figure}

Figure~\ref{fig:plepstar} shows the electron energy and $\shmax$
distributions after cuts have been applied to all variables except the
one being plotted.  The yields and efficiencies are given in
Table~\ref{tab:yields}; these correspond to a branching fraction of
$\left(\VBF\pm \VEstatBF_{(\mathrm{stat})}\right)\times 10^{-3}$ 
using Eq.~\ref{eq:bf}.  We calculate the
partial branching fraction $\Delta\mathcal{B}$ for
$E_e >1.9\,\gev$ in the \B\ rest frame, which corresponds to $E_e>2.0\,\gev$
in the \FourS\ rest frame, and $\shmax(\beta=0)<3.5\,\GEVCC$, unfolded for 
detector effects.  We find 
$\Delta\mathcal{B}(\btouenu)
=\left(\VUBF\pm \VEstatUBF_{(\mathrm{stat})}\right)\times 10^{-4}$.

\begin{table}[!htb]
\caption{Yields and efficiencies.  All uncertainties are statistical
 except for on $N_{\BB}$.}
\begin{center}
\begin{tabular}{|l|c|c|c|c|c|c|} \hline
Data & $N_{\mathrm{cand}}$ & $N_{\mathrm{side}}$ &
 $\epsilon_{\Denu}^{\mathrm{data}} (\%)$ &
 \multicolumn{3}{c|}{$N_{\BB}\ (10^{6})$} \\
 & $8417\pm 164$ & $17776\pm 195$ & $8.72\pm 0.52$
 & \multicolumn{3}{c|}{$88.35\pm 0.97$} \\ \hline
MC   & $M_{\mathrm{bkg}}$  & $M_{\mathrm{side}}$
 & $\epsilon_{\Denu}^{\mathrm{MC}} (\%)$ & $\epsilon_{\,\mathrm{sig}} (\%)$
 & $\epsilon_{\,\overline{\mathrm{sig}}} (\%)$ & $f_u$ \\
 & $5687\pm 47$ & $17904\pm 81$ & $9.18\pm 0.22$ & $3.26\pm 0.03$
 & $0.095\pm 0.003$ & $0.1907\pm 0.0002$ \\ \hline
\end{tabular}
\end{center}
\label{tab:yields}
\end{table}

\section{SYSTEMATIC STUDIES}
\label{sec:Systematics}
Systematic uncertainties are assigned for the modeling of the signal
$\btouenu$ decays and background decays, and the modeling of detector
response.  The leading sources of uncertainty are listed in
Table~\ref{tab:syserr}.  Detector modeling for charged particle
tracking, neutral reconstruction and charged particle identification
was evaluated by comparing data and MC in control samples.  
The simulated production rate of $\KL$ was verified by comparing
data and MC momentum distributions for
$\KS$ in semileptonic decays, and the 
energy deposition of $\KL$ in the calorimeter was varied 
by a generous amount ($\pm 50\%$) to assess the corresponding uncertainty.
The uncertainty due to
Bremsstrahlung in the detector was based on the method used 
in Ref.~\cite{ref:Elmoments}.  
QED final state radiation 
was simulated using PHOTOS~\cite{ref:photos};
comparisons with the analytical result of Ref.~\cite{ref:ginsberg} were
used to assess the systematic uncertainty.  
The uncertainty in the background was
evaluated by varying the form factors and branching fractions of the
$\btocenu$ decays.  An additional uncertainty of $12\%$ was
added to account for the variation in the extracted branching fraction
when the cut on $E_e$ was varied from $1.8\,\gev$ to $2.2\,\gev$.  The
modeling of signal decays is sensitive to the resonant structure at
low mass.  The exclusive branching fractions 
$\mathcal{B}(\bar{B} \rightarrow h e \bar{\nu})$, where 
$h = \pi, \eta, \rho, \omega, \eta^{\prime}$, were varied coherently 
by $\pm 30\%$ to evaluate the uncertainty which proved to be negligeable.
The sensitivity of the branching fraction to the parameters $\mbSF$ and 
$a$ of the De Fazio--Neubert model~\cite{ref:NdF} was evaluated using
uncertainties based on fits~\cite{ref:Gibbons} to the $\b\to\s\gamma$
spectrum~\cite{ref:bsgSpectrum}, leading to
$f_u = 0.190\,^{+0.048}_{-0.024}$.

\begin{table}[!htb]
\caption{Systematic uncertainties on $\mathcal{B}(\btouenu)$ and $\Delta\mathcal{B}(\btouenu)$.}
\begin{center}
\begin{tabular}{|l|c|c|}
\hline
Source of Systematics               & $\delta\mathcal{B}\ (\%)$ &  $\delta(\Delta \mathcal{B})\ (\%)$ \\
\hline
\hline
\hline
1a) Tracking efficiency                 & $\pm 2.7$			& $\pm 2.6 $        \\
\hline
2a) Electron ID efficiency              & $^{+3.1}_{-2.7}$		& $^{+3.1}_{-2.7}$  \\
\hline
3a) Charged particle ID                 & $^{+2.0}_{-2.7}$		& $^{+2.0}_{-2.6}$  \\
\hline
4a) Bremsstrahlung                      & $^{+2.2}_{-1.3}$          & $^{+2.6}_{-0.7}$  \\
\hline
5a) Neutrals reconstruction	       & $\pm 6.3$                 & $\pm 6.4$         \\
\hline
6a) Energy from $K_L^0$   	       & $^{+7.1}_{-7.6}$          & $^{+7.1}_{-7.6}$  \\
\hline
7a) B counting                          & $\pm 1.1$			& $\pm 1.1$	    \\
\hline
\hline
A) Experimental systematics            & $\pm 10.9$ 	        & $\pm 11.0$         \\
\hline
\hline
\hline
1b) $B\rightarrow X_c\ell\nu$ simulation& $^{+6.5}_{-7.5}$          & $\pm{7.5}$        \\
\hline
2b) Radiative corrections               & $^{+4.0}_{-3.1}$	        & $^{+4.5}_{-2.6}$  \\
\hline
3b) Stability scans                    & $\pm 12$                  & $\pm 12$          \\
\hline
\hline
B) Background simulation               & $\pm 14.4$                & $\pm 14.6$        \\
\hline
\hline
\hline
C) Signal simulation                   & $^{+25.4}_{-12.5}$	& $\pm 4.6$         \\
\hline
\hline
\hline
Total (A $\oplus$ B $\oplus$ C)            & $^{+31.2}_{-22.0}$ 	& $\pm 18.8$        \\
\hline
\end{tabular}
\end{center}
\label{tab:syserr}
\end{table}

\section{RESULTS}
\label{sec:Physics}

Using 88~million \FourS\ decays collected with the \babar\ detector
at the \pep2\ \epem\ storage ring, the 
charmless semileptonic branching fraction is determined using
the method of Ref.~\cite{ref:shmax}.
The unfolded partial branching fraction is determined in the \B rest frame for $E_e >1.9\,\gev$ and 
$\shmax(\beta=0)<3.5\,\GEVCC$:
$$\Delta\mathcal{B}(\btouenu)
=\left(\VUBF\pm \VEstatUBF \pm \VEdetUBF \pm \VEbkgUBF
\pm{\VEfuUBFav}\right)\times 10^{-4}~(\mathrm{preliminary}).$$
The uncertainties are from statistics (data and MC), detector
modeling, background modeling and the modeling of \btouenu\ decays, 
respectively. This partial branching fraction can be directly compared with 
theoretical calculations.
We also determine 
$$ \mathcal{B}(\btouenu) = \left(\VBF\pm \VEstatBF \pm \VEdetBF \pm
\VEbkgBF {}^{\VEfuBFp}_{\VEfuBFm}\right)\times 10^{-3}~(\mathrm{preliminary}),$$ 
where the uncertainties are from statistics (data and MC), detector
modeling, background modeling and the SF parameters $\mbSF$
and $a$, respectively.  
The uncertainty due to the
modeling of resonant states in 
$\btouenu$ decays is negligible compared to the
other uncertainties.  

We extract \Vub\ from our measurement using~\cite{ref:BFtoVub}
\begin{displaymath}
\Vub = 0.00424 \sqrt{\frac{\mathcal{B}_{\btouenu}}{0.002}\frac{1.61~{\mathrm{ps}}}{\tau_{B}}}
(1\pm 0.028_{\mathrm{pert}}\pm 0.039_{1/m_{b}^3}),
\label{equation-getvub}
\end{displaymath}
where the first coefficient and the uncertainties quoted in Ref.~\cite{ref:BFtoVub} 
have been updated using the experimental input from the moment measurements obtained 
in Refs.~\cite{ref:Elmoments,ref:Hamoments,ref:OPEfits}. The first coefficient 
takes also into account electroweak radiative corrections~\cite{ref:EWcorrections}.

Taking $\tau_B=1.604\pm 0.012 \ps$ from~\cite{ref:pdg2004} we find
$$
\Vub = (\VVub\pm \VEstatVub\pm \VEdetVub\pm \VEbkgVub ~{}^{\VEfuVubp}_{\VEfuVubm} \pm \VEhqeVub)\times 10^{-3}~(\mathrm{preliminary}),
$$ 
where the first four uncertainties are as for the total branching fraction and the 
last comes from the HQE relating $\Vub$ to the full branching fraction.  

The theoretical uncertainty is dominated by the SF parameter $\mbSF$. 
The variation taken in this analysis is large~\cite{ref:Gibbons}:
$\mbSF=4.735^{+0.110}_{-0.255}\,\gevcc$.  
A more precise estimate of the SF parameters has been determined from the 
$B \rightarrow X_s \gamma$ photon energy spectrum measured by Belle only 
recently~\cite{ref:LimosaniNozaki} from which we obtain the following 
preliminary results:
$$\Delta\mathcal{B}(\btouenu)
=\left(\VUBFBelle\pm \VEstatUBFBelle \pm \VEdetUBFBelle \pm \VEbkgUBFBelle
\pm{\VEfuUBFavBelle}\right)\times 10^{-4},$$
$$ \mathcal{B}(\btouenu) = \left(\VBFBelle\pm \VEstatBFBelle \pm \VEdetBFBelle \pm
\VEbkgBFBelle {}^{\VEfuBFpBelle}_{\VEfuBFmBelle}{}^{\VEbuBFpBelle}_{\VEbuBFmBelle} \right)\times 10^{-3}~(f_u = 0.1628 \pm 0.0002),$$
where the uncertainties are from statistics (data and MC), detector modeling, 
background modeling, the SF parameters $\mbSF$ and $a$, and the uncertainty 
due to the modeling of resonant states in $\btouenu$ decays, respectively, and
$$
\Vub = (\VVubBelle\pm \VEstatVubBelle\pm \VEdetVubBelle\pm \VEbkgVubBelle ~{}^{\VEfuVubpBelle}_{\VEfuVubmBelle} 
~{}^{\VEbuVubpBelle}_{\VEbuVubmBelle} \pm \VEhqeVubBelle) \times 10^{-3},
$$
where the first five uncertainties are as for the total branching fraction and the
last comes from the HQE relating $\Vub$ to the full branching fraction.  
Note that all of the above results are preliminary.

Current theoretical work
(e.g., see Ref.~\cite{ref:BLNP}) is aimed at connecting the $\mbSF$ parameter in
the SF with determinations of $\mbSF$ from moments in $\btocenu$ decays
or from the $\Upsilon(1S)$ mass, and may significantly reduce this
uncertainty.  The results obtained in this analysis 
have very little dependence on the SF parameter $a$ 
and are complementary to
those obtained from studies of the electron endpoint and 
of the mass of the recoiling hadron in semileptonic decays.

\section{ACKNOWLEDGMENTS}
\label{sec:Acknowledgments}


We are grateful for the 
extraordinary contributions of our \pep2\ colleagues in
achieving the excellent luminosity and machine conditions
that have made this work possible.
The success of this project also relies critically on the 
expertise and dedication of the computing organizations that 
support \babar.
The collaborating institutions wish to thank 
SLAC for its support and the kind hospitality extended to them. 
This work is supported by the
US Department of Energy
and National Science Foundation, the
Natural Sciences and Engineering Research Council (Canada),
Institute of High Energy Physics (China), the
Commissariat \`a l'Energie Atomique and
Institut National de Physique Nucl\'eaire et de Physique des Particules
(France), the
Bundesministerium f\"ur Bildung und Forschung and
Deutsche Forschungsgemeinschaft
(Germany), the
Istituto Nazionale di Fisica Nucleare (Italy),
the Foundation for Fundamental Research on Matter (The Netherlands),
the Research Council of Norway, the
Ministry of Science and Technology of the Russian Federation, and the
Particle Physics and Astronomy Research Council (United Kingdom). 
Individuals have received support from 
CONACyT (Mexico),
the A. P. Sloan Foundation, 
the Research Corporation,
and the Alexander von Humboldt Foundation.

\end{document}